\documentclass[aps]{revtex4}
\usepackage{graphicx}
\begin{document}
\title{\bf Spin-magnetophonon level splitting in semimagnetic quantum wells}
\author{ V. L. Gurevich and M. I. Muradov}
\affiliation {A.F.Ioffe Institute, Russian Academy of Sciences,
194021 Saint Petersburg, Russia} \centerline{\null}
\bigskip
\begin{abstract}
Spin-magnetophonon level splitting in a quantum well made of a
semimagnetic wide gap semiconductor is considered. The
semimagnetic semiconductors are characterized by a large effective
$g$ factor. The resonance conditions $\hbar\omega_{\rm
LO}=\mu_BgB$ for the spin flip between two Zeeman levels due to
interaction with longitudinal optical phonons can be achieved
sweeping magnetic field $B$. This condition is studied in quantum
wells. It is shown that it leads to a level splitting that is
dependent on the electron-phonon coupling strength as well as on
the spin-orbit interaction in this structure.

We treat in detail the Rashba model for the spin-orbit interaction
assuming that the quantum well lacks  inversion symmetry and
briefly discuss other models. The resonant transmission and
reflection of light by the well is suggested as a suitable
experimental probe of the level splitting.
\end{abstract} \maketitle
\roman{section}
\section{Introduction}
The resonance coupling of Landau levels with longitudinal optical
phonons (magnetophonon resonance) was theoretically predicted
in~\cite{GF} in  the magnetoresistance investigation. The
resonance takes place every time when the optical phonon frequency
is the cyclotron frequency of an electron times some small
integer. Thus a possibility has been pointed out for an internal
resonance in solids. This phenomenon since its prediction has been
observed in many experiments --- see for instance the
review~\cite{FGPT}.

A possibility of spin-flip transitions of electrons interacting
with optical phonons between the Landau levels of opposite spin
orientations that may be called spin-magnetophonon resonance
(SMPR) --- has been indicated and discussed in a number of papers
(see~\cite{PF,PF3,TAU,Z}). The purpose of the present paper
is to discuss the peculiarities of SMPR in semimagnetic
semiconductors where due to large effective
$g$-factors the corresponding interlevel spacing may be
particularly large and therefore SMPR is well pronounced. The
condition for the spin resonance has the form
\begin{equation}
\label{1i} g\mu_{\rm B}B=\hbar\omega_{\rm LO}.
\end{equation}
Here $\mu_{\rm B}$ is the Bohr magneton, $g$ is the carrier
effective $g$-factor while $B$ is the external magnetic field.

Many remarkable magnetooptical properties of wide gap semimagnetic
semiconductors such as giant exchange splitting of the free
exciton~\cite{KOM}, giant Faraday effect~\cite{KOM,GAJ,BFR}, etc
are determined by a large splitting of conduction and valence
bands in magnetic field. This is a consequence of the exchange
interaction of band carriers with the electrons of the half filled
$d$ shell of the Mn ions. In the present paper we will treat as an
example the compound Cd$_{1-x}$Mn$_{x}$Te where the width of the
gap between the top of  valence band and the bottom of conduction
band in the absence of magnetic field is given by
$E_g=1.595+1.592x\;{\rm eV}$.

In the presence of an external magnetic field the Mn ion spins are
aligned along the magnetic field. Through the exchange of the
Heisenberg type these spins interact with spins of the band
carriers. Eventually, in the mean field model the band carrier
dynamics can be described incorporating  the exchange interaction
only into the enhanced $g$-factor.

There are two competing mechanisms determining the sign and value
of the exchange constant (and of the $g$-factor)~\cite{BHAT, LHEC,
BHAT1}. The first mechanism originating from direct
exchange interaction between the band and $d$ electrons is relatively
weak and ferromagnetic. The second one is due to hybridization of
$d$ orbitals and band states. The latter turns out to be
antiferromagnetic and is negligible for the conduction band while
for the valence band it determines the exchange constant.

The resonance coupling of Landau levels with optical phonons
manifests itself also in a different way though the underlying
physics is the same. It leads to magneto-optical anomalies both in
bulk~\cite{KP} and in two-dimensionally confined
systems~\cite{SM,LKP}. Primary concern of Refs.~\cite{SM,LKP} was
investigation of magneto-optical anomalies of optical phenomena in
conventional GaAs based heterostructures. It was shown that
magneto-optical anomalies in two dimensions provide a powerful
tool for the electron-phonon coupling investigation in these
structures. It was found that under the resonance condition with
respect to electron-phonon interaction the relevant cyclotron peak
splits into a doublet. This effect leads to anomalies in optical
absorption and reflection (as well as in other optical effects
such as for instance Raman scattering).

In the present paper we investigate this effect associated with
SMPR, i. e. the magnetophonon resonances due to the spin flips.
These electron-phonon resonance conditions can occur both for the
valence and  conduction electrons. The exchange constants for the
conduction and valence  electrons turn out to be
different~\cite{GPF}. Though the resonance condition leading to
the level splitting occurs first in the valence band we will show
that the splitting itself is smaller for the valence band states
than for the states in the conduction band.

\begin{figure}[htb]
\begin{center}
  \includegraphics[width=3in]{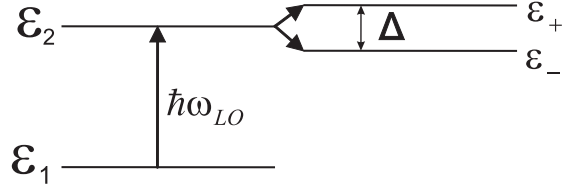}\\
\end{center}
\caption{\label{fig:perv} Level splitting}
\end{figure}

In the next section we consider the level splitting as a formal
quantum mechanical problem. This phenomenon can be understood in
terms of degeneracy lifting of two degenerate states. The energy
degeneracy of an electron state $2$ and an electron in a state $1$
plus an optical phonon~(see Fig.\ref{fig:perv}) is lifted by the
electron-phonon interaction. We will obtain an expression for the
level splitting not specifying the states involved in the relevant
transitions. In  Sec.~\ref{sec:tri} we determine the states and
the energy levels of the conduction electrons taking into account
the spin-orbit interaction in the Rashba model. This allows to
express the level splitting explicitly. We give the required
estimations in the end of this section. As a possible experimental
probe of this splitting phenomenon we propose  the resonant
reflection (transmission) of the light by a quantum well in the
Faraday configuration. We consider the wave reflection
(transmission)  due to direct interband transitions, therefore in
Sec.~\ref{sec:four} we give  explicit expressions for the wave
functions and energies of the valence band states. In
section~\ref{sec:pyat} these wave functions are used to determine
the reflection and transmission coefficient of the light exciting
interband transitions in the quantum well. In Sec.~\ref{val} we
discuss applicability of the perturabation theory for solution of
our problem. We present conclusive remarks in Sec.~\ref{sec:vyv}.
\section{Level splitting}\label{sec:dva}

We begin with treatment of a formal problem: we will consider two
states $1$ and $2$ and find the self energy of an electron in the
state $2$ due to interaction of the electron with optical phonons.
Suppose the energy of the state under consideration
$\varepsilon_{2}$ is close to $\varepsilon_{1}+\hbar\omega_{\rm
LO}$ (i.e. the electron state $2$ and the electron state $1$ plus
the optical phonon with frequency $\omega_{\rm LO}$ are
degenerate). This allows us to put aside all other possible
electron states.

Generally, a single quantum well brings about new phonon
(vibrational) modes. There could be three types of phonons
associated with a quantum well~\cite{MA}: phonons not penetrating
into the quantum well, phonons peaking at the interface and
decaying both in the well and in the barriers (interface phonons),
and phonons confined to the well. The phonon Green function in the
Matsubara technique can be written as
\begin{equation}\label{pgf}
D({\bf r}_{\perp},z,z^{\prime},i\omega_k)=-\sum_{\alpha{\bf
q}_{\perp}}|C_{\alpha}|^2 \left({e^{i{\bf q}_{\perp}{\bf
r}_{\perp}}\eta_{\alpha}(z)\eta_{\alpha}^{\ast}(z^{\prime})\over{i\omega_k+\hbar\omega_{\rm
LO}}}-{e^{-i{\bf q}_{\perp}{\bf
r}_{\perp}}\eta_{\alpha}^{\ast}(z)\eta_{\alpha}(z^{\prime})\over{i\omega_k-\hbar\omega_{\rm
LO}}}\right),
\end{equation}
where $\eta_{\alpha}(z)$ describes spatial distribution of the
phonon $\alpha$ branch in the direction perpendicular to the well
plane ($z$ axis), $\omega_k=2\pi kT$ $(k=0,\pm\,1\ldots)$ are the
Matsubara boson frequencies, $|C_{\alpha}|^2$ is the
electron-phonon coupling strength. $T$ is the temperature; we will
use for it the energy units setting $k_{\rm B}=1$.

The electron-phonon interaction with longitudinal
optical phonons can be treated in the bulk Fr\"{o}hlich~\cite{F}
model. According to the model
$\eta_{\alpha}(z)\rightarrow\,e^{iq_zz}$,
$|C_{\alpha}|^2\rightarrow\,2\pi e^2\hbar\omega_{\rm
LO}/q^2\epsilon^{\ast}$.  This approximation in a relatively wide
wells can be justified noting that the interaction with the
interface phonons in this case can be neglected, interaction with
the confined phonons qualitatively leads to the same result.
Therefore, further on we will work in the Fr\"{o}hlich
approximation. We consider the dispersionless optical phonons,
$\omega_{\rm LO}$ being their frequency and
\begin{equation}
{1\over{\epsilon^{\ast}}}={1\over{\epsilon_{\infty}}}-{1\over{\epsilon_0}},
\end{equation}
where $\epsilon_{\infty}(\epsilon_{0})$ is the high frequency
(static) limit of the dielectric susceptibility.

The electron self energy in the first approximation of the
perturbation theory with respect to the electron-phonon
interaction can be written as (see the diagram (a) in
Fig.(\ref{fig1:diag}))
\begin{equation}\label{se}
\Sigma_{2}(i\varepsilon_n)=-T{2\pi\omega_{\rm
LO}e^2\hbar\over{\epsilon^{\ast}}}\sum_{k}
{F_{21}\over{i(\varepsilon_n-\omega_k)-\varepsilon_1+\mu}}
{2\hbar\omega_{\rm LO}\over{\omega_k^2+(\hbar\omega_{\rm LO})^2}}
\end{equation}
where
\begin{equation}
F_{21}=\int{d^3 q\over{(2\pi)^3}}{\left|<2|e^{i{\bf
qr}}|1>\right|^2\over{q^2}},
\end{equation}
and
\begin{equation}
\varepsilon_n=\pi(2n+1)T.
\end{equation}

To calculate the sum over $k$ in Eq.(\ref{se}) we use
\begin{equation}
G(i\varepsilon_n)={1\over{\pi}}\int_{-\infty}^{\infty}d\varepsilon{{\rm
Im}{G^R(\varepsilon)}\over{\varepsilon-i\varepsilon_n}}
\end{equation}
and
\begin{equation}
D(i\omega_k)={1\over{\pi}}\int_{-\infty}^{\infty}d\varepsilon{{\rm
Im}{D^R(\varepsilon)}\over{\varepsilon-i\omega_k}}.
\end{equation}
We define the phonon Green function $D^R$ as
\begin{equation}
D^R(\omega)={1\over{\omega-\hbar\omega_{\rm LO}+i0}}-{1\over{\omega+\hbar\omega_{\rm LO}+i0}}.
\end{equation}
The sum can be represented as
\begin{equation}
T\sum_kG_{1}(i(\varepsilon_n-\omega_k))D(i\omega_k)=\int_{-\infty}^{\infty}{dxdy\over{\pi^2}}{{\rm
Im} G_{1}^R(x){\rm Im}
D^R(y)\over{x+y-i\varepsilon_n}}[1-n_F(x)+n_B(y)].
\end{equation}
Here we have used the identities
\begin{equation}
T\sum_s{1\over{i\varepsilon_s-x}}=n_F(x)-1,\;\;T\sum_k{1\over{i\omega_k-y}}=-[n_B(y)+1],
\end{equation}
where $n_F(x) [n_B(y)]$ is Fermi [Bose] function. The final result
is
\begin{eqnarray}\label{se111}
\Sigma_{2}(i\varepsilon_n)=-{2\pi\omega_{\rm
LO}e^2\hbar\over{\epsilon^{\ast}}}  F_{21}
\left\{{n_F(\varepsilon_{1})-n_B(\omega_{\rm
LO})-1\over{i\varepsilon_n- (\varepsilon_{1}-\mu)-\hbar\omega_{\rm
LO}}}\right. \left.-{n_F(\varepsilon_{1})+n_B(\omega_{\rm
LO})\over{i\varepsilon_n- (\varepsilon_{1}-\mu)+\hbar\omega_{\rm
LO}}}\right\}.
\end{eqnarray}
Restricting ourselves with the low temperature case
$T\ll\,\hbar\omega_{\rm LO}$ and assuming that the state
$\varepsilon_{1}$ is empty we get
\begin{eqnarray}\label{se1}
\Sigma_{2}(i\varepsilon_n)={\Delta^2/4\over{i\varepsilon_n-
(\varepsilon_{1}-\mu)-\hbar\omega_{\rm LO}}},
\end{eqnarray}
where
\begin{equation}\label{Wdef}
\Delta^2={8\pi\omega_{\rm LO}e^2\hbar\over{\epsilon^{\ast}}}
F_{21}.
\end{equation}
For the electron Green function we get with this self energy
\begin{equation}\label{gf10}
G_{2}(i\varepsilon_n)={1\over{i\varepsilon_n-
\varepsilon_{2}+\mu-(\Delta/2)^2/(i\varepsilon_n-
\varepsilon_{1}-\hbar\omega_{\rm LO}+\mu)}}.
\end{equation}
We perform  analytical continuation replacing
$i\varepsilon_n\rightarrow\,\varepsilon+i0$ and get for the
retarded Green function
\begin{equation}\label{gf12}
G_{2}^R(\varepsilon)={\varepsilon-\varepsilon_{1}-
\hbar\omega_{\rm LO}+\mu\over{(\varepsilon-
\varepsilon_{+}+\mu+i0)(\varepsilon-\varepsilon_{-}+\mu+i0)}}
\end{equation}
where
\begin{equation}\label{eplmin}
\varepsilon_{\pm}={\varepsilon_{2}+
\varepsilon_{1}+\hbar\omega_{\rm LO}\over{2}}\,\pm\,
\sqrt{((\varepsilon_{2}-\varepsilon_{1}- \hbar\omega_{\rm
LO})/2)^2+(\Delta/2)^2}.
\end{equation}
As is seen from Eq.(\ref{gf12}) we have gotten two poles of the
Green function; the level $\varepsilon_{2}$ is split into a
doublet with the energies $\varepsilon_{\pm}$, the spacing between
the poles being equal $\Delta$. The splitting can be expressed
through the parameter $\alpha$ describing the effective mass
polaron shift
\begin{equation}\label{alphashift}
\Delta^2=16\pi\alpha l_{\rm LO}(\hbar\omega_{\rm LO})^2\int{d{\bf
q}\over{(2\pi)^3}}{\left|<2|e^{i{\bf
qr}}|1>\right|^2\over{q^2}},\;\;\alpha^2
={m_ce^4/2(\hbar\epsilon^{\ast})^2\over{\hbar\omega_{\rm LO}}}.
\end{equation}
Here we introduced the length $l_{\rm
LO}=\sqrt{\hbar/2m_c\omega_{\rm LO}}$, $m_c$ is the electron
effective mass. The parameter $\alpha$ for materials with a
relatively weak polarity is small. For instance, $\alpha=0.39$ for
CdTe with partly ionic bonding.
Suppose now that we can achieve  the resonant condition
$\varepsilon_2-\varepsilon_1=\hbar\omega_{\rm LO}$ changing the
interlevel spacing $\varepsilon_2-\varepsilon_1$. If the states
$2$ and $1$ are the spin up and spin down states the resonant
condition can be reached by adjusting external magnetic field.
Since the level splitting is proportional to a matrix element
$<2|e^{i{\bf qr}}|1>$ we see that the phonons can lead to spin
flips only provided the states $2$ and $1$ are not the
eigenfunctions of spin operators ${\bf s}^2$ and $s_z$. For this
reason, we must include into the Hamiltonian the spin-orbit
interaction. We consider the spin-orbit interaction in the Rashba
model~\cite{R}
\begin{equation}\label{rashbaham}
H_R={\alpha_R\over{\hbar}}[{\mbox{\boldmath $\sigma$}}{\bf p}]{\bf
n}.
\end{equation}
Here ${\bf n}$ is a unit vector perpendicular to the quantum well
plane. This interaction is due to the structure inversion
asymmetry. Parameter $\alpha_R$ is of the order of $10^{-9}$
eV$\cdot$cm.

There could be another spin-orbit interaction term that is due to
the bulk inversion asymmetry. The corresponding 3D spin-orbit
Dresselhaus Hamiltonian~\cite{D} in the crystal principal axes
reads
\begin{equation}\label{dressterm}
H_D=\delta({\mbox{\boldmath $\sigma$}}{\bf P}).
\end{equation}
Here $\hbar^3P_x=p_yp_xp_y-p_zp_xp_z$ and other components of
${\bf P}$ can be obtained by cyclic permutations. In 2D case this
Hamiltonian takes the form (we omit the terms cubic in $p$)
\begin{equation}\label{dress}
H_D={\alpha_D\over{\hbar}}(\sigma_yp_y-\sigma_xp_x),
\end{equation}
where $\alpha_D=\delta<p_z^2>/\hbar^2$ and $<p_z^2>$ is averaged
over the transverse motion of the electron.
The parameter $\alpha_D$ can be estimated as $10^{-10}$
eV$\cdot$cm.


Being interested only in the possibility of the line splitting in
the optical reflection (transmission) experiments with quantum
wells explicit calculations  for the spin-orbit interaction of the
Rashba form will be presented since in many semiconductor
nanostructures the Rashba interaction is stronger than the
Dresselhaus one. However, it can be shown that the Dresselhaus
term in the form~(\ref{dress}) does not essentially differ from
the Rashba term, so that  to take into account the Dresselhaus
term one should simply replace the constant $\alpha_R$ by
$\alpha_D$ (this will be sufficient for estimations). Indeed, one
can show that the Dresselhaus term can be obtained from the last
term in Eq.(\ref{ham01}) below simply replacing $\alpha_R$ by $\alpha_D$
and $a$ by $-ia$.
\section{Deep quantum well in transverse magnetic field}\label{sec:tri}
Let $x,y$ be parallel to the quantum well plane, $z$ axis being perpendicular
to the plane of the well. Further on we will consider the simplest
case of an infinitely deep well. We assume that magnetic field $\bf
B$ is along $z$ axis (perpendicular to the plane of the well) and
choose the gauge ${\bf A}=B(0,x,0)$. In  wide gap
semiconductors the conduction and valence bands can be considered
separately. In the zinc blende structures the conduction band
Hamiltonian near the point $\Gamma_6$ is
\begin{equation}\label{ham10}
H=H_0+H_R,
\end{equation}
\begin{equation}\label{ham11}
H_0={\hbar^2\over{2m_c}}\left(-i\nabla+{e\over{\hbar c}}{\bf
A}\right)^2+U+H_Z.
\end{equation}
Here we use the basis of $Ss_{\pm}$ (where $S$ is the $S$-type
Bloch amplitude and $s_-,s_+$ are the two spin functions); $U$ is
the confining potential of the quantum well. We write the Zeeman
Hamiltonian as
\begin{eqnarray}\label{zeeman}
H_Z={1\over{2}}\mu_B\sigma_zg_cB.
\end{eqnarray}
Intending to consider semimagnetic semiconductors
Cd$_{1-x}$Mn$_{x}$Te we will incorporate into the Hamiltonian the
exchange Heisenberg interaction of the conduction band electrons
with Mn ions
\begin{equation}\label{val511}
H_{\rm ce}=-\sum_nJ_{\rm ce}({\bf r}-{\bf R}_n){\bf S}^{\rm Mn}_n
{\bf s},
\end{equation}
where $J_{\rm ce}({\bf r}-{\bf R}_n)$ is the exchange integral of
the electron with the Mn ion localized at ${\bf R}_n$ site, the
sum runs over all the  Mn ions. We will use the mean-field
approximation inserting the mean value of Mn spin in $z$ direction
$<S^{\rm Mn}_z>$ instead of the corresponding operator and
ascribing spin $x<S^{\rm Mn}_z>$ to every crystal site. In
this approximation the exchange Hamiltonian can be rewritten in
the form
\begin{equation}\label{val7}
H_{\rm ce}=-x<S^{\rm Mn}_z>N_0<S|J_{\rm ce}({\bf r})|S> s_z\equiv
-2\hbar\omega_{c}V_cs_z,
\end{equation}
where $N_0$ is the density of unit cells and $<S|J_{\rm ce}({\bf
r})|S>$ is the exchange integral (that is assumed to be positive).
Here for convenience we factor out the cyclotron frequency
$\omega_{c}=eB/m_cc$. The introduced quantity $V_c$ for the
conduction band turns out to be negative and rather large. It can
be written as
$$
V_c=x<S^{\rm Mn}_z>{N_0<S|J_{\rm ce}({\bf
r})|S>\over{2\hbar\omega_{c}}}.
$$
The induced Mn ion spin can be written as
$$
<S^{\rm Mn}_z>=-{\cal B}_S(\zeta),\;\;\zeta={g_{\rm
Mn}\mu_BB\over{k_BT}},
$$
where ${\cal B}_S(\zeta)$ is the Brillouin function
\begin{equation}\label{val7a}
{\cal
B}_S(\zeta)={{2S+1\over{2}}}\coth{\left({2S+1\over{2}}\zeta\right)}
-{1\over{2}}\coth{\left({\zeta\over{2}}\right)}.
\end{equation}
For $S=5/2$
$$
{\cal
B}_{5/2}(\zeta)={35\over{12}}\zeta,\;\;\zeta\,\ll\,1;\;\qquad
{\cal B}_{5/2}(\zeta)={5\over{2}},\;\;\;\zeta\,\gg\,1,
$$
$g_{\rm Mn}=2$, $\mu_B$ is the Bohr magneton, $S=5/2$ is the spin
of a manganese atom. Therefore, we see that $g_c$ in
Eq.~(\ref{zeeman}) must be understood as $g_{zz}-4V_c$. Since
$N_0<S|J_{ce}({\bf r})|S>=0.22$ eV and $\hbar\omega_{c}\sim$
1\,meV we get that $g_c\sim\,50$.

Eigenfunctions of $H_0$ as functions of $y$ can be chosen as plane
waves $e^{ik_yy}/\sqrt{L_y}$. As functions of $z$ they are
the eigenfunctions $\chi_i(z)$ of an infinitely deep one-dimensional well
with associated eigenvalues $\varepsilon_i$. Thus one can rewrite
$H_0$ as
\begin{equation}\label{ham20}
H_0=\varepsilon_i-{\hbar^2\over{2m_c}}{\partial^2\over{\partial
x^2}}+{m_c\omega_c^2\over{2}}(x-x_0)^2+H_Z.
\end{equation}
Here the position of the center of oscillator $x_0=-k_y\hbar c/eB$
depends on the quasimomentum along $y$ direction $\hbar k_y$ (the
motion along $y$ axis is free). The Rashba Hamiltonian in
magnetic field is
\begin{eqnarray}\label{hamr} H_R=\alpha_R
\left(\begin{array}{cc}0&\partial /\partial x+k_y+x/l_c^2\\-\partial /\partial x+k_y+x/l_c^2&0
\end{array}\right).
\end{eqnarray}
Here we have introduced the magnetic length
$l_c=\sqrt{c\hbar/eB}$. Introducing Bose operators according to
$\partial /\partial x=(a-a^{\dagger})/(\sqrt{2}l_c)$,
$x-x_0=l_c(a+a^{\dagger})/(\sqrt{2})$ we get
\begin{eqnarray}\label{ham01}
H=\varepsilon_i+\hbar\omega_c(a^{\dagger}a+1/2)+H_Z+{\sqrt{2}\alpha_R\over{l_c}}
\left(\begin{array}{cc}0&a\\a^{\dagger}&0\end{array}\right).
\end{eqnarray}
The Rashba term does not change the ground state
$\varphi_0(x-x_0)$ and its energy is
$\varepsilon_0=\varepsilon_i+\hbar\omega_c/2-\mu_BBg_{c}/2$. Other
eigenfunctions of $H$ are
\begin{eqnarray}\label{eigfun10}
\psi_{n+}=\left(\begin{array}{c}\cos{u_n}\,\varphi_n(x-x_0)\\\sin{u_n}\,\varphi_{n+1}(x-x_0)\end{array}\right)
\end{eqnarray}
with eigenvalues
\begin{eqnarray}\label{eival10}
\varepsilon_{n+}=\varepsilon_i+\hbar\omega_c(n+1)+
\sqrt{\left({\hbar\omega_c-\mu_Bg_{c}B\over{2}}\right)^2+2{\alpha_R^2\over{l_c^2}}(n+1)}
\end{eqnarray}
and
\begin{eqnarray}\label{eigfun11}
\psi_{n-}=\left(\begin{array}{c}-\sin{u_n}\,\varphi_n(x-x_0)\\\cos{u_n}\,\varphi_{n+1}(x-x_0)\end{array}\right)
\end{eqnarray}
with eigenvalues
\begin{eqnarray}\label{eival11}
\varepsilon_{n-}=\varepsilon_i+\hbar\omega_c(n+1)-\sqrt{\left({\hbar\omega_c-
\mu_Bg_{c}B\over{2}}\right)^2+2{\alpha_R^2\over{l_c^2}}(n+1)},
\end{eqnarray}
where
$$
\tan{2u_n}=2\sqrt{2}{\alpha_R\over{l_c}}{\sqrt{n+1}\over{\mu_Bg_{c}B-\hbar\omega_c}}.
$$
Here $\varphi_{n}$ are the oscillator functions of $x-x_0$
\begin{equation}\label{oscfunctions10}
\varphi_{n}(x-x_0)={1\over{\pi^{1/4}}}{1\over{\sqrt{2^nn!}}}{1\over{l_{c}^{1/2}}}
\exp\left[{-(x-x_0)^2/2l_{c}^2}\right]H_{n}[(x-x_0)/l_{c}]
\end{equation}
where $H_n(x)$ are the Hermite polynomials. Thus we arrive at two
groups of levels 
separated by (large) energy $\mu_Bg_cB$.
Both groups consist of sublevels that are nearly equidistant (if
one neglects the spin-orbit contribution to the energy) separated
by the cyclotron energy $\hbar\omega_c$. The minimal energy in the
first group is $\varepsilon_0$, in the second group it is
$\varepsilon_{0+}=\varepsilon_0+\mu_Bg_cB$.

\begin{figure}[htb]
\begin{center}
\includegraphics[width=3in]{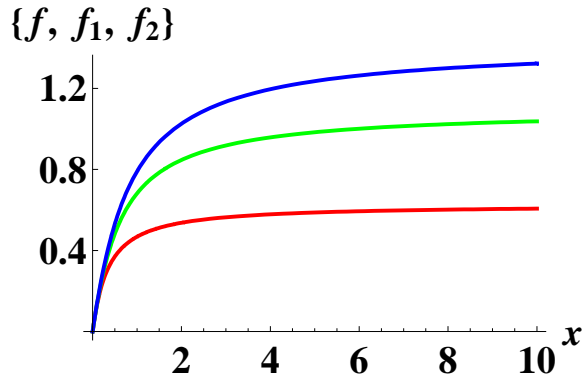}\\
\end{center}
\caption{\label{fig:trii} Functions $f(x)$, $f_{0}(x)$  and
$f_{1}(x)$. The second
 function corresponds to $\Delta^2_{1+,0-}$ and saturates at $x\gg\,1$ reaching the value $7\sqrt{2\pi}/16$
while the third one corresponds to the transition $1-\rightarrow\,2+$ and also saturates at $x\gg\,1$
reaching the value $145\sqrt{2\pi}/256$}.
\end{figure}

In what follows we restrict ourselves with the phonon induced
transitions $0\rightarrow\,0+$ and $n-\rightarrow\,(n+1)+$. As we
will see below under the realistic conditions the estimations show
that the level splitting due to electron-phonon interaction turns
out to be small as compared to the cyclotron energy, therefore it
is sufficient to consider each pair of states separately. Indeed,
for typical magnetic fields of the order of several tesla the
magnetic length $l_c=\sqrt{\hbar c/eB}\,\sim\,10$ nm, and the
cyclotron energy $\hbar\omega_{c}=\hbar^2/m_cl_c^2\,\sim\,10^{-2}$
eV while the level splitting is of the order
$\hbar\Delta\omega\sim\,10^{-3}$ eV (see the estimations at the
end of this section).
Let us now calculate the matrix element between the ground state
$0$ of the first group and the $0+$ state of the second one
\begin{equation}\label{me10}
|<i,0+,k_y|e^{iq_xx+iq_zz}|i,0,k_y-q_y>|^2=\sin^2{u_0}|<i|e^{iq_zz}|i>|^2
{l_c^2q_\perp^2\over{2}}\exp{\left(-{l_c^2q_\perp^2\over{2}}\right)}.
\end{equation}
or taking into account that $u_0\,\ll\,1$
\begin{equation}\label{alphashift10}
\Delta^2_{0+,0}=4{\alpha}{l_{\rm LO}\over{l_c}}(\hbar\omega_{\rm
LO})^2u_0^2f\left({l_c\over{L}}\right)
\end{equation}
where
\begin{equation}
f\left({l_c\over{L}}\right)=l_c\int\,dz_1dz_2\chi_i^2(z_1)\chi_i^2(z_2)\int_0^{\infty}dq_{\perp}
{q_{\perp}^2l_c^2\over{2}}\exp{\left(-l_c^2q_{\perp}^2/2-q_{\perp}|z_1-z_2|\right)}.
\end{equation}
Transitions from $n-$ to $(n+1)+$ states are resonant too; for
these transitions we have (omitting $k_y$ and $k_y-q_y$)
\begin{eqnarray*}\label{me101}
|<i,(n+1)+|e^{iq_xx+iq_zz}|i,n->|^2=
2\left({\alpha_R\over{l_c\mu_Bg_cB}}\right)^2|<i|e^{iq_zz}|i>|^2
\\\times\,{l_c^2q_\perp^2\over{2}}e^{-l_c^2q_\perp^2/2}
\left[L_{n+1}^1(l_c^2q_\perp^2/2)-
L_{n}^1(l_c^2q_\perp^2/2)\right]^2,
\end{eqnarray*}
where $L_n^{\alpha}(x)$ are the Laguerre polynomials defined as in
Ref.\cite{GR}.
Here we have taken into account that
$\sin{u_n}\simeq\,u_n\simeq\,\sqrt{2(n+1)}\alpha_R/l_c\mu_BB$.
Since
$L_{n+1}^{\alpha-1}(x)=L_{n+1}^{\alpha}(x)-L_{n}^{\alpha}(x)$ this
expression can be simplified
\begin{eqnarray*}\label{me102}
|<i,(n+1)+|e^{iq_xx+iq_zz}|i,n->|^2=
2\left({\alpha_R\over{l_c\mu_Bg_cB}}\right)^2|<i|e^{iq_zz}|i>|^2
{l_c^2q_\perp^2\over{2}}e^{-l_c^2q_\perp^2/2}
\left[L_{n+1}(l_c^2q_\perp^2/2)\right]^2.
\end{eqnarray*}
We have
\begin{equation}
\Delta^2_{(n+1)+,n-}=8{\alpha}{l_{\rm
LO}\over{l_c}}(\hbar\omega_{\rm LO})^2
\left({\alpha_R\over{l_c\mu_Bg_cB}}\right)^2f_{n}\left({l_c\over{L}}\right),
\end{equation}
\begin{equation}
f_{n}\left({l_c\over{L}}\right)=l_c\int\,dz_1dz_2\chi_i^2(z_1)\chi_i^2(z_2)\int_0^{\infty}dq_{\perp}
{l_c^2q_{\perp}^2\over{2}}\left[L_{n+1}\left({l_c^2q_\perp^2\over{2}}\right)
\right]^2e^{-l_c^2q_{\perp}^2/2-q_{\perp}|z_1-z_2|}
\end{equation}
Here we give an estimation for $\Delta^2_{0+,0}$ assuming that the
transverse motion is described by the wave function
$\chi_1(z)=\sqrt{2/L}\sin{\left(\pi z/L\right)}$ [see
Eq.(\ref{alphashift10})]
\begin{equation}\label{estalphashift10}
\Delta^2_{0+,0}=8{\alpha}(\hbar\omega_{\rm
LO})^2\left(\alpha_R\over{l_c\mu_Bg_cB}\right)^2 {l_{\rm
LO}\over{l_c}}f\left({l_c\over{L}}\right),
\end{equation}
where $f(x)=\sqrt{2\pi}/4$ for $x\,\gg\,1$ and $f(x)=3x/2$ for
$x\,\ll\,1$ (see Fig.\ref{fig:trii}).
Taking into account the CdTe parameters, namely the longitudinal
optical phonon frequency $\omega_{\rm LO}=3.22\cdot\,10^{13}$
s$^{-1}$ ($246$ K), the susceptibilities $\epsilon_0=10.3$ and
$\epsilon_{\infty}=6.9$, the effective electron mass $m_c=0.1m_0$
($m_0$ is the free electron mass) we see that the line splitting
is $\Delta\omega=\Delta/\hbar\simeq\,\alpha_R\sqrt{\alpha}/\hbar
l_c\sim\,10^{11}$ s$^{-1}$. For the Dresselhaus interaction the
line splitting would be $\alpha_D/\alpha_R$ times smaller.
\section{Valence band}\label{sec:four}
In the  structures having zinc blende symmetry the valence $\Gamma_8$ band
is described by the Luttinger Hamiltonian
\begin{equation}\label{val1}
H=H_0+H(k_z),
\end{equation}
where we separate the part $H(k_z)$ depending on $k_z$,
\begin{equation}\label{val2}
H_0=-{\hbar^2\over{2m_0}}\left\{(\gamma_1+{5\over{2}}\gamma_2)(
k_x^2+k_y^2)-2\gamma_2(J_x^2k_x^2+J_y^2k_y^2)-4\gamma_3\{J_xJ_y\}
\{k_xk_y\}+2{e\over{c}}\kappa{\bf JB} \right\},
\end{equation}
\begin{equation}\label{val3}
H(k_z)=-{\hbar^2\over{2m_0}}\left\{(\gamma_1+{5\over{2}}\gamma_2)k_z^2-2\gamma_2J_z^2
k_z^2-4\gamma_3(\{J_xJ_z\}\{k_xk_z\}+\{J_yJ_z\}\{k_yk_z\})\right\}.
\end{equation}
Here $\gamma_1,\gamma_2,\gamma_3,\kappa$ are material parameters,
${\bf J}$ is the operator of angular momentum $J=3/2$, the
symmetrized products are defined according to
\begin{equation}\label{val4}
\{AB\}={AB+BA\over{2}}.
\end{equation}
We add to the valence band Hamiltonian the exchange Heisenberg
interaction of the valence band electrons with Mn ions
\begin{equation}\label{val5}
H_{ve}=-\sum_nJ_{ve}({\bf r}-{\bf R}_n){\bf S}^{\rm Mn}_n
{\bf s}
\end{equation}
where $J_{ve}({\bf r}-{\bf R}_n)$ is the exchange integral of a
valence band electron with a Mn ion.

The wave function can be written as
\begin{equation}\label{val6}
\Psi=\sum_iF_i({\bf r})u_i({\bf r}),
\end{equation}
where $u_i(\bf r)$ are the four degenerate states at the top of
the valence band~\cite{BP}
\begin{equation}\label{val6a}
u_{\pm\,3/2}=\mp{1\over{\sqrt{2}}}(X\pm iY)s_{\pm},\;\;
u_{\pm\,1/2}={1\over{\sqrt{3}}}\left[\mp{1\over{\sqrt{2}}}(X\pm
iY)s_{\mp}+ \sqrt{2}Zs_{\pm}\right].
\end{equation}

It is easily seen that in this basis the spin operator $s_z=\sigma_z/2$ is also diagonal and is related to the
$J_z$ operator by $s_z=J_z/3$, therefore we can rewrite the
exchange Hamiltonian as
\begin{equation}\label{val7b}
H_{e}=-x<S^{\rm Mn}_z>N_0<X|J_{ve}({\bf r})|X>
{1\over{3}}J_z\equiv -2\hbar\omega_{c0}V_vJ_z.
\end{equation}
Here for convenience we factor out the cyclotron frequency
$\omega_{c0}=eB/m_0c$ anticipating its appearance in the following
formulae. The introduced quantity $V_v$ for the valence band turns
out to be positive and rather large. It can be estimated as
$$
V_v=x<S^{\rm Mn}_z>{N_0<X|J_{ve}({\bf
r})|X>\over{6\hbar\omega_{c0}}}.
$$
Here the exchange integral for the valence band $<X|J_{ve}({\bf
r})|X>$  is negative.

At the typical magnetic fields  of the order of several tesla the
magnetic length $l_c=\sqrt{\hbar c/eB}\,\sim\,10$ nm, and the
cyclotron energy
$\hbar\omega_{c0}=\hbar^2/m_0l_c^2\,\sim\,10^{-3}$ eV, while
$N_0<X|J_{ve}({\bf r})|X>=-0.88$ eV. Therefore $V_v\,\gg\,1$.

We again choose the gauge  ${\bf A}=B(0,x,0)$ and introduce the
operators $a,a^{\dagger}$ according to
\begin{equation}\label{val8}
k_x=-{i\over{\sqrt{2}l_c}}(a-a^{\dagger}),\;k_y={1\over{\sqrt{2}l_c}}(a+a^{\dagger}).
\end{equation}
Replacing also the operators $J_x,J_y$ by $J_{\pm}=J_x\pm\,iJ_y$
we get
\begin{equation}\label{val9}
H_0=-\hbar\omega_{c0}\left\{[\gamma_1-{5\over{4}}\gamma_2+
\gamma_2J_z^2](a^{\dagger}a+1/2)+
{\gamma_2\over{4}}(J_{-}^2+J_{+}^2)[a^2+(a^{\dagger})^2]+
{\gamma_3\over{4}}(J_{+}^2-J_{-}^2)[a^2-(a^{\dagger})^2]+{e\over{c}}l_c^2\kappa
\bf{}JB\right\},
\end{equation}
\begin{eqnarray}\label{val100}
H(k_z)=-\hbar\omega_{c0}\left\{
aJ_{+}(J_z+{1\over2})-a^{\dagger}J_{-}(J_z-1/2)\right\}i\sqrt{2}
\gamma_3(l_ck_z)
-\hbar\omega_{c0}{1\over{2}}\left\{\gamma_1+{5\over{2}}
\gamma_2-2\gamma_2J_z^2\right\}(l_ck_z)^2. \end{eqnarray} Further
on we will use the spherical approximation, i.e. we set
$\gamma_2=\gamma_3$. We
get
\begin{eqnarray}\label{val10}
-{H\over{\hbar\omega_{c0}}}=2g_v
J_z+(\gamma_1-{5\over{4}}\gamma_2+\gamma_2J_z^2)
(a^{\dagger}a+1/2)+{1\over{2}}
\left\{\gamma_1+{5\over{2}}\gamma_2-2\gamma_2J_z^2
\right\}(l_ck_z)^2\\\nonumber+
{\gamma_2\over{2}}[J_{+}^2a^2+J_{-}^2(a^{\dagger})^2]+\left\{
aJ_{+}(J_z+1/2)-a^{\dagger}J_{-}(J_z-1/2)\right\}i\sqrt{2}
\gamma_2(l_ck_z)
\end{eqnarray}
where we have introduced the effective $g$ factor in the valence
band $g_v={\hbar\kappa/{2}}+V_v$.

Due to large values of the exchange Hamiltonian $g_v$ we can omit
the last two terms, i.e.
\begin{equation}\label{val111}
V=\left\{
aJ_{+}(J_z+1/2)-a^{\dagger}J_{-}(J_z-1/2)\right\}\sqrt{2}
\gamma_2l_c{\partial\over{\partial z}}+
{\gamma_2\over{2}}(J_{+}^2a^2+J_{-}^2(a^{\dagger})^2).
\end{equation}
in Eq.~(\ref{val10}) that sufficiently simplifies the problem. The
reason of such a separation of the Hamiltonian is rather obvious,
the Hamiltonian $V$ leads to transitions changing both spin and
Landau numbers and can be taken into account as a perturbation. In
this approximation the levels can be considered independently and
we have for the top heavy and light hole series of levels (in the
hole representation)
\begin{equation}\label{envbhh}
E_{-3/2,n,n_v}^{(\rm
hh)}=E_g-3\hbar\omega_{c0}g_v+\hbar\omega_{c0}{m_0(3m_h+m_l)
\over{4m_lm_h}}(n+1/2)+{\pi^2\hbar^2n_v^2\over{2m_hL^2}}
\end{equation}
\begin{equation}\label{wfvbhh}
\psi_{-3/2}^{(\rm
hh)}=\varphi_n(x-x_{0k_y})\chi_{n_v}(z){e^{ik_yy}\over
{\sqrt{L_y}}}u_{-3/2}
\end{equation}
\begin{equation}\label{envblighth}
E_{-1/2,n,n_v}^{(\rm
lh)}=E_g-\hbar\omega_{c0}g_v+\hbar\omega_{c0}{m_0(3m_l+m_h)
\over{4m_lm_h}}(n+1/2)+{\pi^2\hbar^2n_v^2\over{2m_lL^2}}
\end{equation}
\begin{equation}\label{wfvblighth}
\psi_{-1/2}^{(\rm
lh)}=\varphi_n(x-x_{0k_y})\chi_{n_v}(z){e^{ik_yy}\over{\sqrt{L_y}}}u_{-1/2}
\end{equation}
Here $E_g$ is the gap, $m_{l}(m_h)$ are the light (heavy) hole
masses, $n_v$ is the  quantization number of transverse motion and
$\chi_{n_v}(z)$ is the corresponding wave function. We take into
account that the $\gamma_1,\gamma_2$ parameters are related to
effective masses by $\gamma_1=m_0(m_{h}+m_{ l})/2m_{h}m_{l}$ and
$\gamma_2=m_0(m_{ h}-m_{l})/4m_{ h}m_{ l}$.

In this zeroth approximation phonons can not induce transitions
between these states. In the next approximation of perturbation
theory with respect to $V$ these states are mixed and we get for
the top heavy hole state $\psi_{-3/2,n,n_v}^{(\rm hh)}$
\begin{equation}\label{wfvbhhfa}
\psi_{-3/2,0,1}^{(\rm
hh)}=\varphi_0(x-x_{0k_y})\chi_{1}(z){e^{ik_yy}\over
{\sqrt{L_y}}}u_{-3/2}.
\end{equation}

For the light hole top state we have
\begin{equation}\label{wfvbhhfa2}
\psi_{-1/2,0,1}^{(\rm lh)}={e^{ik_yy}\over
{\sqrt{L_y}}}\left\{\varphi_0(x-x_{0k_y})\chi_{1}(z)
u_{-1/2}+{4\over{3}}{\gamma_2\hbar\omega_{c0}(l_c/L)\over{E_{-1/2,0,1}-E_{-3/2,1,2}}}
\varphi_1(x-x_{0k_y})\chi_{2}(z)u_{-3/2}\right\}.
\end{equation}
Now it is obvious that phonon can induce transitions between these
states. Suppose that sweeping the magnetic field we achieve the
hole-phonon resonance condition between the states described by
Eq.(\ref{wfvbhhfa}) and Eq.(\ref{wfvbhhfa2})
$$
E_{-1/2,0,1}-E_{-3/2,0,1}=\hbar\omega_{\rm LO}
$$
or
$$
2g_v\hbar\omega_{c0}-{m_0(m_{ h}-m_{l})\over{4m_{ h}m_{
l}}}\left(\hbar\omega_{c0}-2{\pi^2\hbar^2\over{m_0L^2}}\right)=\hbar\omega_{\rm
LO}.
$$
For the value $\Delta_{-1/2,-3/2}\equiv\,\Delta_v$ describing the
splitting in the valence band we get at the resonant condition
$$
\Delta^2_{v}=4\alpha{l_{\rm LO}\over{l_c}}(\hbar\omega_{\rm
LO})^26\left({m_0(m_{ h}-m_{ l})\over{4m_{h}m_{
l}}}\right)^2\left(\hbar\omega_{c0}\over{\hbar\omega_{\rm
LO}}\right)^2\left(l_c\over{L}\right)^2f_v(l_c/L),
$$
where
\begin{equation}
f_{v}\left({l_c\over{L}}\right)=l_c\int\,dz_1dz_2\chi_1(z_1)\chi_2(z_1)
\chi_1(z_2)\chi_2(z_2)\int_0^{\infty}dq_{\perp}
{q_{\perp}^2l_c^2\over{2}}\exp{\left(-l_c^2q_{\perp}^2/2-q_{\perp}|z_1-z_2|\right)}.
\end{equation}
$f_v(x)=x,\;\;x\,\ll\,1,\;\;f_v(x)=(10/9\pi^2x),\;\;x\,\gg\,1$.

Let us compare the SMPR splittings in the conduction and valence
bands. We evaluate
$$
{\Delta_{0+,0}\over{\Delta_{v}}}\sim\,\left({\alpha_R\over{L\hbar^2/2m_0L^2}}\right)
\left({g_v\over{g_c}}\right)^{3/4} \left({4m_{h}m_{
l}\over{m_0(m_{h}-m_{
l})}}\right)\left({f(l_c/L)\over{f_v(2\sqrt{g_v/g_c}l_c/L)}}\right)^{1/2}
$$
and see that the splitting in the conduction band is bigger than
in the valence band and is determined by the parameter $g_v/g_c$.
Here $l_c$ is the magnetic length for magnetic fields required to
achieve the resonance condition in the conduction band.

In principle, in valence band one can also write the spin-orbital
term of Rashba type~\cite{Win}
$$
H_{vR}={\alpha^{'}\over{\hbar}}[{\bf J p}]{\bf n},
$$
that in magnetic field can be rewritten as
$$
H_{vR}={\alpha^{'}\over{\sqrt{2}l_c}}\left\{J_+a+J_-a^{+}\right\}.
$$
This term leads to the ratio
$$
{\Delta_{0+,0}\over{\Delta_{v}}}\sim\,\left({\alpha_R\over{\alpha^{'}}}\right)
\left({g_v\over{g_c}}\right)^{3/4}
\left({f(l_c/L)\over{f_v^{\prime}(2\sqrt{g_v/g_c}l_c/L)}}\right)^{1/2},
$$
where
\begin{equation}
f_{v}^{\prime}\left({l_c\over{L}}\right)=\int\,dz_1dz_2\chi_1^2(z_1)
\chi_2^2(z_2)\int_0^{\infty}dq
{q^4}\exp{\left(-q^2/2-q|z_1-z_2|/l_c\right)}.
\end{equation}
Although the SMPR condition are met first for the hole states as
one sweeps the magnetic field the splitting in the valence band
turns out to be much smaller than in the conduction band. This is
the consequence of the smaller spin-phonon coupling strength for
the states strongly shifted by the Zeeman energy.

\section{Resonant reflection and transmission}\label{sec:pyat}
We consider the simplest geometry where the wave $\sim\,e^{ikz}$
is incident perpendicularly to the plane of the well. Neglecting
in the induced current the longitudinal part (this term in the
induced current has a small factor
$u_0\simeq\,\alpha_R/\mu_Bg_cBl_c$ ) so that we can put
$\nabla\cdot{\bf D}=\epsilon_b\nabla\cdot{\bf E}=0$ the
Maxwell equation for the wave with frequency $\omega$ can be
written as (in this section $k$ denotes the wave vector of light)
\begin{equation}\label{me11}
{d^2\over{dz^2}}E_{\alpha}+k^2E_{\alpha}={4\pi\over{\hbar
c^2}}\int\,dz^{\prime}\Pi_{\alpha\beta}^R(z,z^{\prime},\omega)E_{\beta}(z^{\prime}).
\end{equation}
Here $k^2=\omega^2\epsilon_b/c^2$ (we neglect the difference in
the background susceptibilities of the well and barriers). We have
taken into account that the polarization operator
$\Pi_{\alpha\beta}$ (here the averaging over the distances much
greater than the lattice parameter is implied) is
\begin{equation}\label{polarizoper}
\Pi^R(z,z^{\prime},\omega)=\int\,dx^{\prime}dy^{\prime}\Pi^R({\bf
r},{\bf r}^{\prime},\omega).
\end{equation}
The Green function of operator $d^2/dz^2+k^2$ obeys the equation
\begin{equation}\label{grfun}
\left({d^2\over{dz^2}}+k^2\right){\cal G}(z,z^{\prime})=-\delta
(z-z^{\prime})
\end{equation}
and is given by
\begin{equation}
{\cal
G}^{\pm}(z,z^{\prime})=\pm{i\over{2k}}e^{\pm\,ik|z-z^{\prime}|}.
\end{equation}
For the transmission and reflection problem one should  use ${\cal
G}^{+}(z,z^{\prime})$ function. Then the solution of
Eq.~(\ref{me11}) can be written as
\begin{equation}\label{sol}
E_{\alpha}=E^{0}_{\alpha}e^{ikz}-{4\pi\over{\hbar
c^2}}\int\,dz^{\prime}dz^{\prime\prime}{\cal G}^{+}(z,z^{\prime})
\Pi_{\alpha\beta}^R(z^{\prime},z^{\prime\prime},\omega)E_{\beta}(z^{\prime\prime}).
\end{equation}
where $E^{0}_{\alpha}$ is the amplitude of the incident wave. For
$z>L$, where $L$ is the width of the quantum well, we can identify
the transmitted wave as
\begin{equation}\label{sol1}
E^{t}_{\alpha}=E^{0}_{\alpha}e^{ikz}-{2i\pi\over{k\hbar
c^2}}e^{ikz}\int_0^L\,dz^{\prime}dz^{\prime\prime}e^{-ikz^{\prime}}
\Pi_{\alpha\beta}^R(z^{\prime},z^{\prime\prime},\omega)E_{\beta}(z^{\prime\prime})
\end{equation}
and the reflected one can be identified considering $z<0$
\begin{equation}\label{sol2}
E_{\alpha}^r=-{2i\pi\over{k\hbar
c^2}}e^{-ikz}\int_0^L\,dz^{\prime}dz^{\prime\prime}e^{ikz^{\prime}}
\Pi_{\alpha\beta}^R(z^{\prime},z^{\prime\prime},\omega)E_{\beta}(z^{\prime\prime}).
\end{equation}
Assuming that $\Pi^R(z,z^{\prime})$ can be factorized as
$\Pi^R(z,z^{\prime})=\Pi^{(1)}(z)\Pi^{(2)}(z^{\prime})$ (such a
factorization is possible since below we will consider transitions
between two fixed states with respect to transverse motion
$\chi_{n_v}(z)$,$\chi_{n_c}(z)$) we scalarly multiply
Eq.~(\ref{sol}) by $\Pi_{\alpha}^{(2)}(z)$ and integrate over $z$,
then we get
\begin{equation}\label{sol3}
{\cal F}=-{\cal F}{4\pi\over{\hbar c^2}}\int\,dz\,dz^{\prime}{\cal
G}^{+}(z,z^{\prime}) \Pi_{\alpha\alpha}^R(z,z^{\prime},\omega)+
E^{0}_{\alpha}\int_0^L\,dz\,e^{ikz}\Pi^{(2)}_{\alpha}(z),
\end{equation}
where we have introduced notation
$$
{\cal F}=\int_0^L\,dz^{\prime}
\Pi^{(2)}_{\beta}(z^{\prime},\omega)E_{\beta}(z^{\prime}).
$$
Solving Eq.~(\ref{sol3}) for ${\cal F}$ and making use of
Eq.~(\ref{sol1}) and Eq.~(\ref{sol2}) we get for the amplitudes of
the transmitted and reflected waves
\begin{equation}\label{sol11}
E^{t}_{\alpha}=\left(\delta_{\alpha\beta}+
{4\pi\int\,dz^{\prime}\,dz\,e^{-ik(z-z^{\prime})}\Pi_{\alpha\beta}^R(z,z^{\prime},\omega)
\over{2ikc^2\hbar-4\pi\int\,dz^{\prime}\,dz\,e^{ik|z-z^{\prime}|}
\Pi_{\gamma\gamma}^R(z,z^{\prime},\omega)}}\right)E^{0}_{\beta},
\end{equation}
\begin{equation}\label{sol21}
E^{r}_{\alpha}=
{4\pi\int\,dz^{\prime}\,dz\,e^{ik(z+z^{\prime})}\Pi_{\alpha\beta}^R(z,z^{\prime},\omega)
\over{2ikc^2\hbar-4\pi\int\,dz^{\prime}\,dz\,e^{ik|z-z^{\prime}|}
\Pi_{\gamma\gamma}^R(z,z^{\prime},\omega)}}E^{0}_{\beta}.
\end{equation}
In the basis ${\bf e}_{\pm}=({\bf e}_x\pm\,i{\bf e}_y)/\sqrt{2}$
in our approximation only one component of $\Pi_{\alpha\beta}$ is
nonvanishing, i.e. $\Pi_{++}=2\Pi_{xx}$. Due to the symmetry
relations we have $\Pi_{xx}=\Pi_{yy}=i\Pi_{xy}=-i\Pi_{yx}$.
Therefore, left circularly polarized incident wave ${\bf e}_{-}$
is not reflected, while for the right polarized incident wave
${\bf e}_{+}$ we get
\begin{equation}\label{cir01}
t_+=1+{4\pi\int\,dz^{\prime}\,dz\,e^{-ik(z-z^{\prime})}\Pi_{++}^R(z,z^{\prime},\omega)
\over{2ikc^2\hbar-4\pi\int\,dz^{\prime}\,dz\,e^{ik|z-z^{\prime}|}
\Pi_{++}^R(z,z^{\prime},\omega)}}
\end{equation}
for the transmission coefficient ${\bf E}^t=t_+E^0_+{\bf
e}_+e^{ikz}$ and
\begin{equation}\label{cir02}
r_+={4\pi\int\,dz^{\prime}\,dz\,e^{ik(z+z^{\prime})}\Pi_{++}^R(z,z^{\prime},\omega)
\over{2ikc^2\hbar-4\pi\int\,dz^{\prime}\,dz\,e^{ik|z-z^{\prime}|}
\Pi_{++}^R(z,z^{\prime},\omega)}}
\end{equation}
for the reflection (amplitude) coefficient ${\bf E}^r=r_+E^0_+{\bf
e}_+e^{-ikz}$. Since the propagation direction of the wave is now
inverted the reflected wave has the left polarization. A linearly
polarized incident wave will be reflected as a circularly left
polarized wave. In the case where the wave length $2\pi/k$ is
bigger than the well width $L$ (i.e. $kL\,\ll\,1$) the exponential
factors can be omitted.

Let us consider the polarization operator. We can write the formal
expression for the operator
\begin{eqnarray}\label{polop10}
\Pi^R_{\alpha\beta}({\bf r},{\bf
r}^{\prime},\omega)=-{i\over{2}}\sum_{\lambda_1\lambda_2}j^{\alpha}_{\lambda_2\lambda_1}({\bf
r})j^{\beta}_{\lambda_1\lambda_2}({\bf
r}^{\prime})\int\,{d\varepsilon\over{2\pi\hbar}}\left\{
\tanh{{\varepsilon+\hbar\omega}\over{2T}}\left(G^R_{\lambda_1}(\varepsilon/\hbar+\omega)-
G^A_{\lambda_1}(\varepsilon/\hbar+\omega)\right)\right.
G^{A}_{\lambda_2}(\varepsilon/\hbar)\\\nonumber
\left.+\tanh{{\varepsilon}\over{2T}}\left(G^R_{\lambda_2}(\varepsilon/\hbar)-
G^A_{\lambda_2}(\varepsilon/\hbar)\right)
G^{R}_{\lambda_1}(\varepsilon/\hbar) \right\},
\end{eqnarray}
where
\begin{equation}\label{curr}
{\bf j}_{\lambda_1\lambda_2}({\bf
r})={ie\hbar\over{2m_0}}\left\{\Phi^{\ast}_{\lambda_1}({\bf
r})\nabla\,\Phi_{\lambda_2}(\bf
r)-\left(\nabla\,\Phi^{\ast}_{\lambda_1}(\bf
r)\right)\Phi_{\lambda_2}(\bf r)\right\}-{e^2\over{m_0c}}{\bf
A}_0\Phi^{\ast}_{\lambda_1}(\bf r)\Phi_{\lambda_2}(\bf r).
\end{equation}
Here $\Phi_{\lambda}(\bf r)$ are the eigenfunction of the
Hamiltonian and ${\bf A}_0$ is the vector potential of the applied
static magnetic field. We consider the interband transitions and
assume that the valence band states are occupied while the states
in the conduction band are empty. Keeping only the resonant
contribution in Eq.(\ref{polop10}) we get
\begin{equation}\label{polop11}
\Pi^R_{\alpha\beta}({\bf r},{\bf
r}^{\prime},\omega)=\sum_{\lambda_c\lambda_v}j^{\alpha}_{\lambda_v\lambda_c}({\bf
r})j^{\beta}_{\lambda_c\lambda_v}({\bf
r}^{\prime})\int\,{d\varepsilon\over{2\pi\hbar\,i}}G^R_{\lambda_c}
(\varepsilon/\hbar+\omega)G^{A}_{\lambda_v}(\varepsilon/\hbar),
\end{equation}
The states in the valence band we can consider as unchanged by the
electron-phonon interaction (since we are interested only in the
splitting phenomenon in the conduction band) and we get for the
circularly polarized wave
\begin{equation}\label{refcof}
r_+={-i\Gamma(\hbar\omega-\varepsilon_{1}-\hbar\omega_{\rm
LO}+\varepsilon_v)
\over{(\hbar\omega-\varepsilon_++\varepsilon_v+i0)
(\hbar\omega-\varepsilon_-+\varepsilon_v+i0)/\hbar+i\Gamma(\hbar\omega-
\varepsilon_{1}-\hbar\omega_{\rm LO}+\varepsilon_v)}},
\end{equation}
where we have introduced notation
\begin{equation}\label{fdef}
\Gamma={4\pi\,\over{\hbar\omega
c\sqrt{\epsilon_b}}}\sum_{\lambda_c\lambda_v}\int\,d{\bf
r}^{\prime}dz\,j^x_{\lambda_c\lambda_v}({\bf
r}^{\prime})j^x_{\lambda_v\lambda_c}({\bf r}).
\end{equation}
This quantity can be related to the  recombination rate of
the transition under consideration. Here we have taken into
account that $\Pi_{xx}=\Pi_{yy}$.
\begin{figure}[htb]
\begin{center}
  \includegraphics[width=3in]{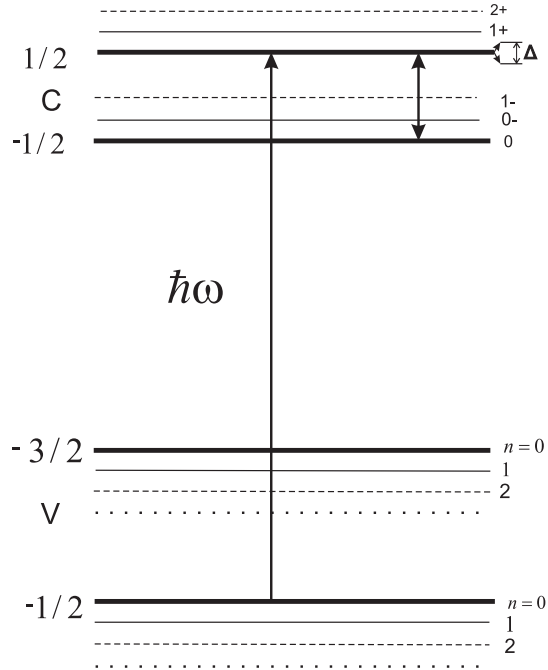}\\
\end{center}
\caption{\label{fig:predposl}Interband transitions.}
\end{figure}
Since we are interested in the splitting phenomenon we assume that
the resonance light frequency is close to the transition from the
ground state in the $+$ group in the conduction band to the ground
state in the valence band with $J_z=-1/2$ (see
Fig.\ref{fig:predposl})~\cite{F1}. Therefore, in the following
formulae we set
$\varepsilon_{1}=\varepsilon_{c1}=\varepsilon_{0}$,
$\varepsilon_{2}=\varepsilon_{c2}=\varepsilon_{0+}$,
$\varepsilon_{c2}-\varepsilon_{c1}=\mu_Bg_cB$,
$\varepsilon_{v}=-E_{-1/2,0,1}^{(\rm lh)}$ and
$\Delta=\Delta_{0+,0}$.

The specification of the transition between the states described
by Eq.(\ref{wfvblighth}) and
\begin{equation}\label{wfcondb}
\psi_{c0+}=S\chi_{n_c}(z)\left(\varphi_0(x-x_{0k_y})s_{+}+u_0
\varphi_{1}(x-x_{0k_y})s_{-}\right){e^{ik_yy}\over{\sqrt{L_y}}}
\end{equation}
allows us to express the recombination rate explicitly
\begin{equation}\label{fdef1}
\Gamma={4\pi\,\over{\hbar\omega c\sqrt{\epsilon_b}}}
{e^2|p_{cv}|^2\over{6m_0^2}}{1\over{2\pi
l_c^2}},\;\;p_{cv}=<S|p_x|X>.
\end{equation}
Here we have assumed $|<\chi_{v1}|\chi_{c1}>|^2=1$ for the
overlapping of the transverse quantized wave function of the
conduction and valence bands. Now we will introduce the following
dimensionless variables: the deviation from the  SMPR
$(\varepsilon_{c2}-\varepsilon_{c1}-\hbar\omega_{\rm
LO})/\Delta=\delta$, the optical frequency
$2\hbar(\omega-\omega_0)/\Delta=x$ [
$\omega_0=(\varepsilon_{c2}-\varepsilon_{v})/\hbar$ being the
interband resonance frequency], and the uncertainty in the level
energy position $2\hbar\,\Gamma/\Delta=\gamma$. Then we can write
the power reflection coefficient $R=|r_+|^2$ as
\begin{equation}\label{refcof1}
R={\gamma^2(x+2\delta)^2\over{(x+\delta-\sqrt{1+\delta^2})^2
(x+\delta+\sqrt{1+\delta^2})^2+\gamma^2(x+2\delta)^2}}.
\end{equation}
If the dimensionless deviation $\delta\,\gg\,1$ (i.e. the
deviation from the phonon resonance condition is much bigger than
the splitting) using $\sqrt{1+\delta^2}\simeq\,\delta$ we see that
the single line structure is restored
\begin{equation}\label{refcof2}
R={\gamma^2\over{x^2 +\gamma^2}}.
\end{equation}
In the case of exact electron phonon resonance  that can be
achieved by sweeping the magnetic field, $\delta=0$ and  we get
\begin{equation}\label{refcof22}
R={\gamma^2x^2\over{(x^2-1)^2+\gamma^2x^2}}.\end{equation} In this
case the power reflection coefficients reaches its maximal value
under the optical resonant conditions. For linearly polarized
incident wave this maximal value is $1/2$.

So far we have assumed that the energy uncertainty of the level
under consideration is much smaller than the splitting $\Delta$,
otherwise the level splitting can not be resolved. Indeed, we will
have  for the Green function instead of Eq.(\ref{gf10})
\begin{equation}\label{disgf10}
G_{2}(\varepsilon)={1\over{\varepsilon+i\hbar\Gamma_{2}-\varepsilon_{2}+\mu-(\Delta/2)^2
/(\varepsilon+i\hbar\Gamma_{1}-\varepsilon_{1}-\hbar\omega_{\rm
LO}+\mu)}}
\end{equation}
provided we take this uncertainty into account. Here we have
phenomenologically introduced $\Gamma_{2}$ and $\Gamma_{1}$ for
the corresponding energy levels $\varepsilon_{2}$ and
$\varepsilon_{1}$.
 It is seen
from this expression that even for
$\varepsilon-\varepsilon_{1}-\hbar\omega_{\rm LO}=0$ we can
discard the second term in the denominator since
$\Delta\,\ll\,\hbar\Gamma_{1}$ and the level does not split.
\begin{figure}[htb]
\begin{center}
  \includegraphics[width=4in]{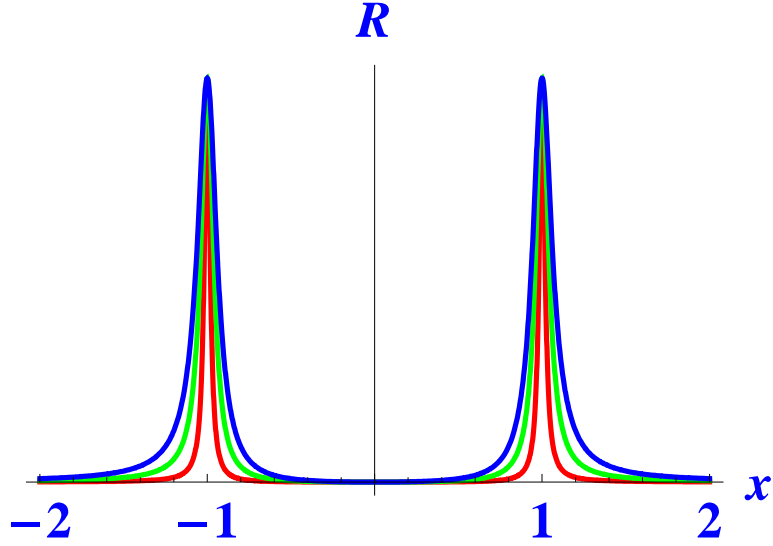}\\
\end{center}
\caption{\label{fig:posl} Power reflection coefficient for SMPR as
a function of optical frequency $x$ for $\gamma=0.04,0.09,0.14$.
Here only the natural level width (due to recombination processes)
is taken into account.}
\end{figure}
The recombination rate can be estimated taking into account that
$|p_{cv}|^2/2m_0$ is of the order of the Bohr energy,
$\hbar\omega\sim\,E_g\sim\,1.6$ eV. Then it is seen that
$\gamma=\hbar\Gamma/\Delta\,\ll\,1$ as is assumed in
Fig.\ref{fig:posl}.

\begin{figure}[htb]
\begin{center}
  \includegraphics[width=3.6 in]{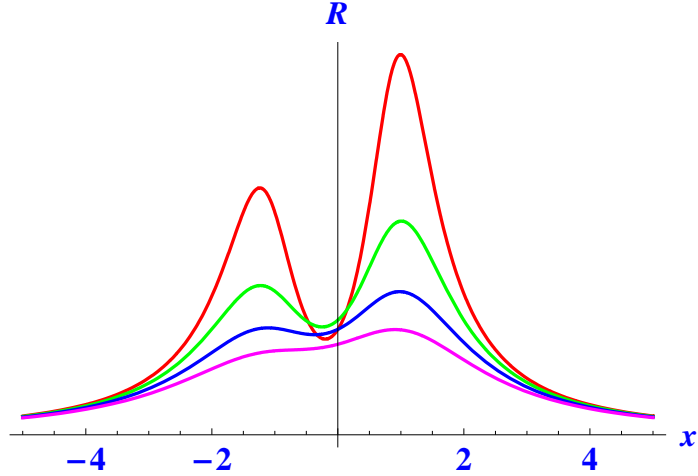}\\
\end{center}
\caption{\label{fig:tret} Reflection coefficient as a function of
optical frequency at $\delta=0.1$ for different level widths
$\gamma_e=0.7,1,1.3,1.6$.}
\end{figure}

\begin{figure}[htb]
\begin{center}
  \includegraphics[width=4 in]{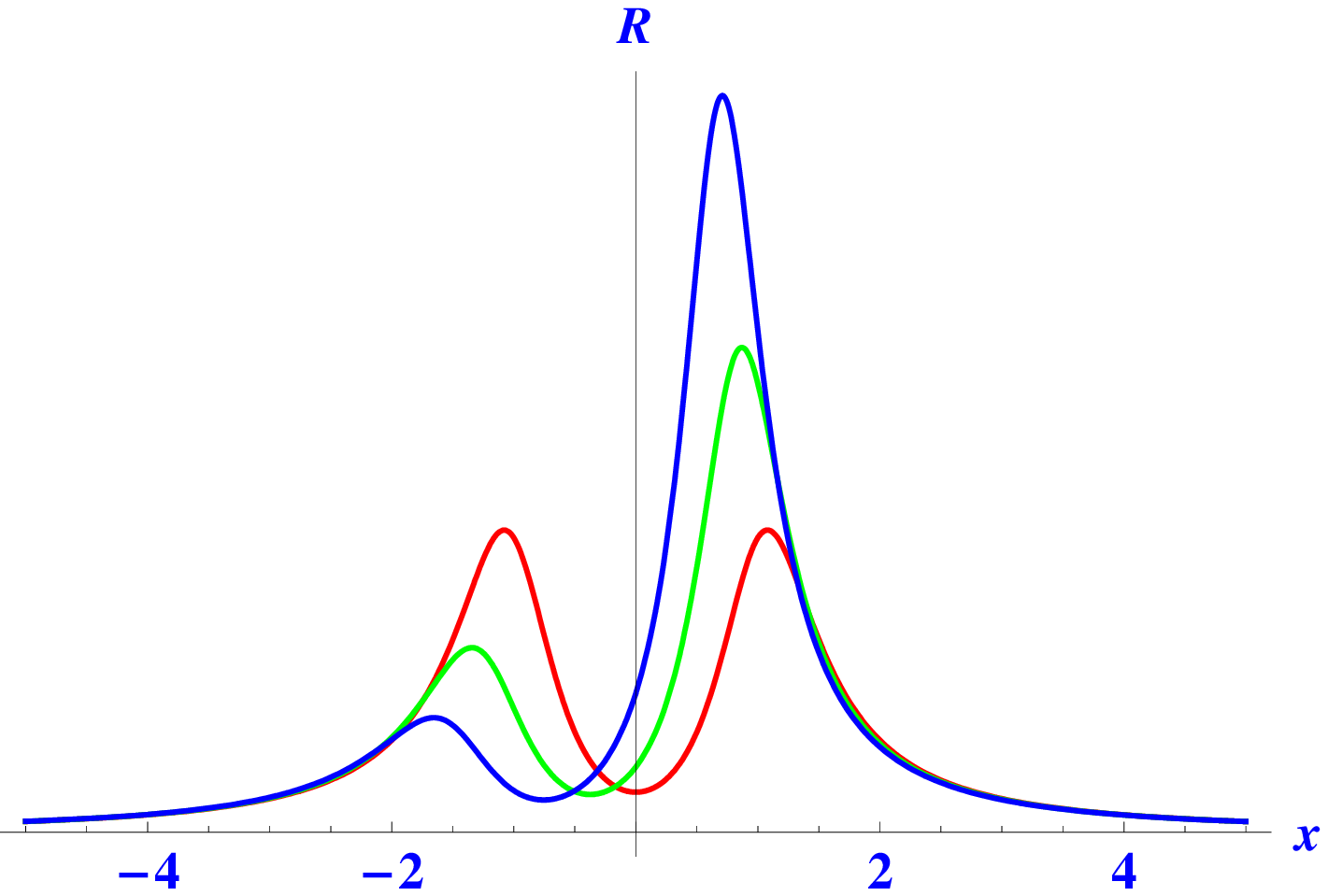}\\
\end{center}
\caption{\label{fig:vtor} Reflection coefficient for different
deviations from the spin electron-phonon resonance
$\delta=0,0.2,0.4$ at $\gamma_e=0.5$.}
\end{figure}

Let us consider the case of equal $\Gamma_1=\Gamma_2$ widths of
both levels. Then we can write for the reflection coefficient
\begin{equation}\label{coeff}
R={\gamma^2[(x+2\delta)^2+\gamma_e^2]\over{(x+\delta-\sqrt{1+\delta^2+\gamma_e^2})^2
(x+\delta+\sqrt{1+\delta^2+\gamma_e^2})^2+4\gamma_e^2(x+\delta)^2}},
\end{equation}
where we introduce a dimensionless quantity proportional to the
sum of level widths $\gamma_e=4\hbar\Gamma_{1}/\Delta$ and neglect
the level width due to recombination processes.
Fig.~\ref{fig:tret} demonstrates how the increasing of the level
widths smears the doublet structure of the reflection line. The
symmetry of this doublet structure depends on the deviation from
the spin electron-phonon resonance (Fig.~\ref{fig:vtor}).

\section{Applicability of perturbation theory}\label{val}
In Sec.\ref{sec:dva} we considered only the simplest diagram for
the self energy. Now we are going to discuss the validity of this
approximation for the spin-phonon interaction. It is easily seen
that each additional phonon line in the higher order diagrams can
bring about additional resonant denominator, therefore we should
consider the series of the most diverging sequence of diagrams.
The situation is not unique and has been encountered earlier in
the polaron problem in the three dimensional case and first such a
consequence of diagrams has been considered by
Pitaevskii~\cite{Pit}.

We consider two empty states $1$ and $2$ with energies
$\varepsilon_{1,2}$ $\varepsilon_2=\varepsilon_1+\hbar\omega_{\rm
LO}$. Each state is unoccupied $\varepsilon_{1,2}\,>\,\mu$.
Therefore we can write for the electron Green's function
\begin{equation}\label{egf}
G(\varepsilon,{\bf r}_1,{\bf r}_2)=\sum_{\nu=1}^2{\Psi_\nu({\bf
r}_1)\Psi^{\ast}({\bf
r}_2)\over{\varepsilon-(\varepsilon_{\nu}-\mu)+i0}}.
\end{equation}
The phonon Green function can be written as
\begin{equation}\label{phongreenf}
D(\omega,{\bf r}_1,{\bf r}_2)=\sum_{\alpha{\bf q}}|C_{\alpha,{\bf
q}}|^2 \left({{e^{-i{\bf q}({\bf r}_1-{\bf
r}_2)}}\over{\omega-\hbar\omega_{\rm LO}+i0}}-{e^{i{\bf q}({\bf
r}_1-{\bf r}_2)}\over{\omega+\hbar\omega_{\rm LO}}-i0}\right),
\end{equation}
where $|C_{\alpha}|^2\rightarrow\,2\pi e^2\hbar\omega_{\rm
LO}/q^2\epsilon^{\ast}$. We are to evaluate the Green function for
the state $2$. Since we consider the empty electron states above
the chemical potential the self energy diagrams will involve Green
functions of the type
$$
{1\over{\varepsilon-\omega-(\varepsilon_{1,2}-\mu)+i0}}.
$$
These functions have the pole with respect to $\omega$ in the
upper half-plane. Therefore we keep in the phonon Green's function
only the part having the pole with respect to $\omega$ in the
lower half-plane (otherwise the integration over $\omega$
vanishes), i.e.
$$
{1\over{\omega-\omega_{\rm LO}+i0}}.
$$

\begin{figure}[htb]
\begin{center}
  \includegraphics[width=5in]{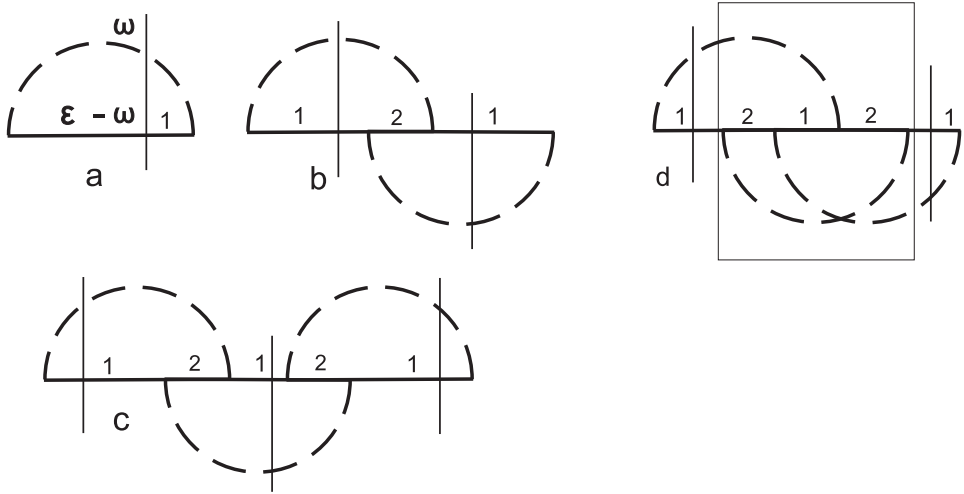}\\
\end{center}
\caption{\label{fig1:diag} Self energy diagrams. Resonant sections
are shown by vertical lines}
\end{figure}

\begin{figure}[htb]
\begin{center}
  \includegraphics[width=5in]{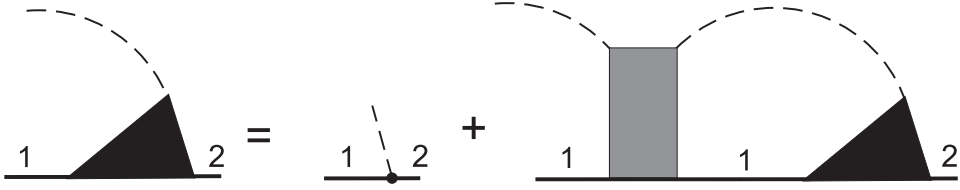}\\
\end{center}
\caption{\label{fig2:diag} Equation for the vertex}
\end{figure}

\begin{figure}[htb]
\begin{center}
  \includegraphics[width=5in]{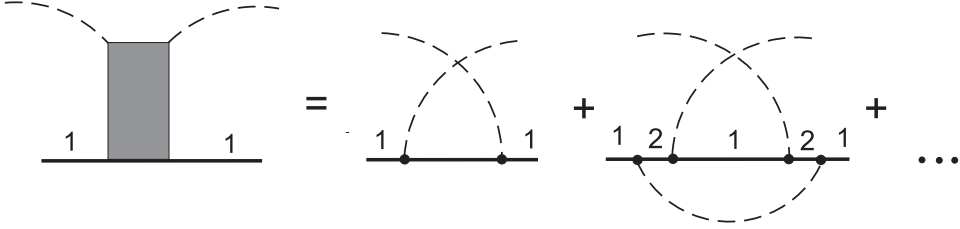}\\
\end{center}
\caption{\label{fig3:diag} Block expansion}
\end{figure}

The simplest electron self energy diagram (see diagram (a) in
Fig.\ref{fig1:diag}) has a resonant denominator
\begin{equation}\label{sediagr}
\Sigma_{2}(\varepsilon)\sim\,i\int{d\omega\over{2\pi}}G(\varepsilon-\omega)D(\omega)=
 {1\over{\varepsilon-\omega_{\rm LO}-(\varepsilon_1-\mu)+i0}},
\end{equation}
when $\varepsilon$ is in the vicinity of
$\varepsilon_2=\varepsilon_1+\omega_{\rm LO}$.
Diagrams with more resonances are of two types: the first type
leads to corrections  to the Green function (to the line $1$ in
the skeleton diagram (a) in Fig.\ref{fig1:diag})) and they can be
taken into account regarding the Green function as renormalized,
the second type leads to the corrections to the electron-phonon
vertex. Since the corrections of the first type can be taken into
account perturbatively (these diagrams do not involve resonant
denominators) we will not consider them  and  concentrate on the
diagrams of the second type. Several diagrams of the last type are
presented in Fig.\ref{fig1:diag}. The diagrams $(b)$ and $(c)$
involve two and three resonant denominators, respectively. We can
draw more complicated diagrams with two resonance denominators
(similar to diagram (d) in Fig.\ref{fig1:diag})), it is now seen
that the diagrams of this type can be regarded as the diagram
$(b)$ with a block that does not involve resonant denominators,
such a block we will call a compact block. Therefore, we can write
the integral equation for the renormalized vertex (see
Fig.\ref{fig2:diag}). In the Fig.\ref{fig3:diag} we show that the
compact block is the expansion with respect to electron-phonon
coupling strength, therefore we write the integral equation
keeping only the first term in this expansion
\begin{eqnarray}\label{ievp}
\Gamma(\varepsilon-\omega,k_y-q_y,\varepsilon,k_y,q_x,q_z)=<1k_y-q_y|e^{-iq_xx-iq_zz}|2k_y>+\\
+i\int{d\omega^{\prime}d{\bf q}^{\prime}\over{(2\pi)^4}}
<1k_y-q_y|e^{iq_x^{\prime}x+iq_z^{\prime}z}|2k_y-q_y-q_y^{\prime}>
<2k_y-q_y-q_y^{\prime}|e^{-iq_xx-iq_zz}|1k_y-q_y^{\prime}>\times\\
\times{|C_{{\bf
q}^{\prime}}|^2\Gamma(\varepsilon-\omega^{\prime},k_y-q_y^{\prime},\varepsilon,k_y,q_x^{\prime},
q_z^{\prime})\over{(\omega^{\prime}-\omega_{\rm LO}+i0)
(\varepsilon-\omega-\omega^{\prime}-(\varepsilon_2-\mu)+i0)
(\varepsilon-\omega^{\prime}-(\varepsilon_1-\mu)+i0)}}.
\end{eqnarray}
Let us write this equation for the specific states $1=\{n=0,i=1\}$
and $2=\{n=0+,i=1\}$ (see Sec.\ref{sec:tri}), i.e. we consider the
ground states with respect to orbital motion and  to spatial
confinement. In order to simplify the integral equation we
introduce the function $A(q_{\perp}, \varepsilon,\omega)$ by
relation
\begin{equation}\label{iesimp}
\Gamma(\varepsilon-\omega,k_y-q_y,\varepsilon,k_y,q_x,q_z)=e^{iq_x(k_y-q_y/2)}e^{-q^2_{\perp}/4}
{q_y-iq_x\over{\sqrt{2}}}<1|e^{-iq_zz}|1>A(q_{\perp},
\varepsilon,\omega),
\end{equation}
where $q_{\perp}=\sqrt{q_x^2+q_y^2}$ and wave vectors are
dimensionless (the factor is the magnetic length). Then, using for
the phase factors under the integral the relation
\begin{equation}\label{ierel}
e^{i(q_yq_x^{\prime}-q_xq_y^{\prime})}=\sum_{n=-\infty}^{\infty}J_2(q_{\perp}q_{\perp}^{\prime})
e^{-in(\varphi^{\prime}-\varphi)}
\end{equation}
we get
\begin{eqnarray}\label{iered}
A(q, \varepsilon,\omega)=1+{i\over{2}}
\int{d\omega^{\prime}p^3dp\over{(2\pi)^3}} e^{-p^2/2}{J_2(q
p)\phi(p)
A(p,\varepsilon,\omega^{\prime})\over{(\omega^{\prime}-\omega_{\rm
LO}+i0)
(\varepsilon-\omega-\omega^{\prime}-(\varepsilon_2-\mu)+i0)
(\varepsilon-\omega^{\prime}-(\varepsilon_1-\mu)+i0)}}.
\end{eqnarray}
\begin{eqnarray}\label{iereduc}
\phi(p)={u_0^2\over{l_c}}\int\,dq_z|C_{p,q_z}|^2|<1|e^{iq_zz/l_c}|1>|^2.
\end{eqnarray}
Now we suppose that the function $A(q, \varepsilon,\omega)$ has no
poles with respect to $\omega$ in the lower half complex plane and
consider the case when the magnetic length is much bigger than the
quantum well width. The last assumption leads to
$\phi(p)=\pi/pl_c$ and we can rewrite the integral equation for
$A(q, \varepsilon,\omega_{\rm LO})\equiv\,A(q,\varepsilon)$ as the
Fredholm equation
\begin{eqnarray}\label{iefredholm}
A(q, \varepsilon)=1+\lambda \int_0^{\infty}\,dp\, p^2
e^{-p^2/2}J_2(q p) A(p,\varepsilon),
\end{eqnarray}
where  parameter $\lambda$ includes the resonant denominator
\begin{equation}\label{iela}
\lambda=-{1\over{8}}{u_0^2(e^2/\varepsilon^{\ast}l_c)\over{\varepsilon-(\varepsilon_1+\hbar\omega_{\rm
LO}-\mu)+i0}}.
\end{equation}
In reality the uncertainty of the level $i\hbar/\tau$ enter the
last equation instead of $i0$. Let us evaluate the minimum of
$\tau$ when we remain in the framework of perturbation theory and
it is then sufficient to consider only the skeleton diagram for
the self energy. With $\hbar\omega_{\rm LO}=0.02$ eV, $B=3$ T,
$m_c=0.1m_0$, $\alpha_R=10^{-9}$ eV$\cdot$cm, $\alpha=0.39$ we get
that the perturbation scheme is valid for the relaxation times
shorter than $\tau_0=5\cdot\,10^{-10}$ sec. On the other hand, to
resolve the splitting the level uncertainty must be smaller than
the level splitting $\Delta\simeq\,5\cdot\,10^{-4}$ eV. This
requires the times bigger than $10^{-12}$ sec. Therefore, there
exists a region of relaxation times $\sim\,10^{-11}$ sec., where
the perturbation theory is valid and the splitting phenomena is
discernable.

Here we wish to note that in the ordinary situation of the
magnetophonon phenomena one meets the opposite case of big
$\lambda$ parameter in Eq.(\ref{iefredholm}) and the integral
equation should be solved. Therefore, we suppose that the theories
taking into account only the one phonon processes described by the
skeleton self energy diagram can not be considered as reliable.
\section{Conclusive remarks}\label{sec:vyv}

We have considered optical manifestation of SMPR in semimagnetic
semiconductors. Due to the electron-phonon coupling the resonant
reflection and transmission line representing the interband
transitions is split  into two lines. The distance between the
lines is determined by the strength of electron-phonon coupling.

We should, however, indicate that some points have not been taken
into account in our calculation. Among them the most important is
the natural width of the phonon levels. For the optical phonons at
low temperatures it is determined by decay of an optical phonon
into two acoustic ones.

The natural width of the electron lines is also important. It may
be determined by  collisions of electrons with acoustic phonons
and with  defects of the lattice, as well as by recombination.
These effects result in widening of the lines that has been
briefly discussed. Under the conditions where these effects are
strong, the lines may overlap as has been indicated above.

So far we have considered a situation where the equilibrium
concentration of the carriers is so low that they do not influence
the light absorption. One can conceive, however, another case of
interest where, for instance, in equilibrium electrons (provided
by donors outside the well) fill the conduction band up to the
Fermi level. In such a case transitions between the valence band
and the states of the conduction band above the Fermi level are
allowed. The oscillator strength for these transitions may be
bigger than for those treated in the present paper. One can expect
that the width of the electron level in the conduction band should
be rather small as the electrons can emit acoustic phonons with
the energies not bigger than the spacing between the level they
occupy and the Fermi level. However, one can expect that the width
of the level in the valence band may be much bigger. Indeed, the
holes can emit phonons with comparatively large energies as the
spacing between their level and the top of the valence band can be
rather large.

Experimental observation of SMPR can provide information about the
electron-phonon interaction. Its investigation can also provide
important information concerning various contributions into
spin-orbit interaction as well as the strength of the exchange
interaction.

\acknowledgments

The authors are grateful to Yu. G. Kusrayev for an interesting discussion where
the topic of the present investigation emerged.


\begin{thebibliography}{99}
\bibitem{GF} V. L. Gurevich, Yu. A. Firsov, Zh. Eksp. Teor. Fiz.
{\bf 40}, 199, 1961 [Sov. Phys.-JETP {\bf 13}, 137]
\bibitem{FGPT} Yu. A. Firsov, V. L. Gurevich, R. V. Parfeniev, I.
M. Tsidil'kovskii, in {\em Landau Level Spectroscopy} edited by G.
Landwehr and E. I. Rashba (Elsevier, Amsterdam, 1991)
\bibitem{PF} Pavlov S. T., Yu. A. Firsov, Fiz. Tverd. Tela {\bf 7}, 2634, 1965[Sov. Phys.-Solid State {\bf 7}, 2131,1966],
Zh. Eksp. Teor. Fiz. {\bf 49}, 1664, 1965 [Sov. Phys.-JETP {\bf
22}, 1137,1966], Fiz. Tverd. Tela {\bf 9}, 1780, 1967 [Sov.
Phys.-Solid State {\bf 9}, 1394]
\bibitem{PF3} S. T. Pavlov and Yu. A. Firsov, Fiz. Tverd. Tela
{\bf9}, 1780 (1967) [Sov. Phys. - Solid State {\bf9}, 1394
(1967)].

\bibitem{TAU} I. M. Tsidilkovskii, M. M. Aksel'rod, and S. I.
Uritskii, Phys. Status Solidi {\bf12}, 667 (1965).

\bibitem{Z}W. Zawadzki, G. Bauer, and H. Kahlert, \prl {\bf35},
1098 (1975).
\bibitem{KOM} Komarov A. V., Ryabchenko S. M., Terletskii O. V., Zheru I. I., Ivanchuk R. D., Zh. Eksp. Teor. Fiz.
{\bf 73}, 608, 1977 [Sov. Phys.-JETP {\bf 46}, 318]
\bibitem{GAJ} Gaj J. L., Galazka R. R., Nawrocki M., Solid. State
Commun., {\bf 25}, 193, 1978
\bibitem{BFR} Bartholomew D. U., Furdyna J. K., Ramdas A. K.,
Phys. Rev. B {\bf 34}, 6943, 1986
\bibitem{BHAT} A. K. Bhattacharjee, Fishman G., Coqblin B.,
Physica 117B and 118B, 449, 1983
\bibitem{LHEC} B. E. Larsen, K. C. Hass, H. Ehrenreich, A. E.
Carlsson, Phys. Rev. B {\bf 37}, 4137, 1988
\bibitem{BHAT1} Bhattacharjee A. K., Phys. Rev. B, {\bf 41}, 5696, 1990
\bibitem{KP} L. I. Korovin, S. T. Pavlov, Zh. Eksp. Teor. Fiz.
{\bf 53}, 1708, 1967 [Sov. Phys.-JETP {\bf 26}, 979, 1968]
\bibitem{SM} S. Das Sarma, A. Madhukar, Phys. Rev. B, {\bf 22},
2823, 1980

\bibitem{LKP} I. G. Lang, L. I. Korovin, S. T. Pavlov, Fiz. Tverd. Tela, {\bf
47},1704, 2005)[Phys.-Solid State,{\bf 47}, 2005]; Fiz. Tverd.
Tela, {\bf 48},1693, 2006 [Phys.-Solid State,{\bf 48}, 2006]

\bibitem{GPF} J. A. Gaj, R. Planel, G. Fishman, Solid State
Commun., {\bf 29}, 435, 1979
\bibitem{MA} N. Mori, T. Ando, Phys. Rev. B, {\bf 40}, 6175, 1989
\bibitem{F} H. Fr\"{o}hlich, Adv. Phys., {\bf 3},2, 325, 1954
\bibitem{R} Bychkov Yu. L., Rashba E. I., Pisma Zh. Eksp. Teor.
Fiz., {\bf 39}, 66, 1984 [Sov. Phys.-JETP {\bf 39}, 78, 1984]
\bibitem{D} G. Dresselhaus, Phys. Rev., {\bf 100}, 580, 1955
\bibitem{GR} Gradshteyn I. S., I. M. Rhyzhik, Tables of integrals,
series and products (Academic Press, New York, 1965)
\bibitem{BP} Bir G. L., Pikus G. E., Symmetry and Strain-Induced Effects in Semiconductors, (Wiley, New
York, 1974)

\bibitem{Win} R. Winkler, Phys. Rev. B, {\bf 62}, 4245, 2000
\bibitem{F1} Note that the spin-orbit interaction can allow optical transitions from this group to the valence band top states with $J_z=-3/2$ but the intensity of these transitions is proportional to the small interaction constant $u_0$.
\bibitem{Pit} L. P. Pitaevskii, Zh. Eksp. Teor. Fiz. {\bf 36}, 1168, 1959 [Sov. Phys. -JETP {\bf 19}, 1959]
\end{thebibliography}
\end{document}